\documentclass[final,5p,times,twocolumn]{elsarticle}
\biboptions{sort&compress}
\usepackage{amsfonts}
\usepackage{amssymb}
\usepackage{subfig}
\usepackage{float}
\usepackage{xcolor}
\usepackage{soul}

\usepackage{amsmath}
\usepackage{graphicx}
\usepackage{color}
\usepackage{mathtools,leftidx}
\usepackage{eso-pic}
\DeclareMathAlphabet\mathbfcal{OMS}{cmsy}{b}{n}

\makeatletter
\def\ps@pprintTitle{%
   \let\@oddhead\@empty
   \let\@evenhead\@empty
   \let\@oddfoot\@empty
   \let\@evenfoot\@oddfoot
}
\makeatother

\begin{document}
\AddToShipoutPictureBG*{%
  \AtPageUpperLeft{%
    \hspace{0.94\paperwidth}%
    \raisebox{-3\baselineskip}{%
      \makebox[0pt][r]{\tt IFT-UAM/CSIC-20-60}
}}}%
\begin{frontmatter}
\title{A non-perturbative approach to the scalar Casimir effect with Lorentz symmetry violation}

\author[1]{C. A. Escobar}
\ead{carlos$\_$escobar@fisica.unam.mx}

\author[2]{A. Mart\'{i}n-Ruiz}
\ead{alberto.martin@nucleares.unam.mx}

\author[2]{O. J. Franca}
\ead{francamentesantiago@ciencias.unam.mx}

\author[3]{Marcos A. G. Garcia}
\ead{marcosa.garcia@uam.es}

\address[1]{Instituto de F\'{i}sica, Universidad Nacional Aut\'{o}noma de M\'{e}xico, Apartado Postal 20-364, Ciudad de M\'{e}xico 01000, M\'{e}xico}
\address[2]{Instituto de Ciencias Nucleares, Universidad Nacional Aut{\'o}noma de M{\'e}xico, A. Postal 70-543, 04510 Ciudad de M\'{e}xico, M{\'e}xico}
\address[3]{Instituto de F\'isica Te\'orica (IFT) UAM-CSIC, Campus de Cantoblanco, 28049 Madrid, Spain}

\begin{abstract}

We determine the effect of Lorentz invariance violation in the vacuum energy and stress between two parallel plates separated by a distance $L$, in the presence of a massive real scalar field. We parametrize the Lorentz-violation in terms of a symmetric tensor $h^{\,\mu\nu}$ that represents a constant background. Through the Green's function method, we obtain the global Casimir energy, the Casimir force between the plates and the energy density in a closed analytical form without resorting to perturbative methods. With regards to the pressure, we find that $\mathcal{F}_c(L)=\mathcal{F}_0(\tilde{L})/\sqrt{-{\rm det}\, h^{\,\mu\nu}}$, where $\mathcal{F}_0$ is the Lorentz-invariant expression, and $\tilde{L}$ is the plate separation rescaled by the component of $h^{\,\mu\nu}$ normal to the plates, $\tilde{L}=L/\sqrt{-h^{nn}}$. We also analyze the Casimir stress including finite-temperature corrections. The local behavior of the Casimir energy density is also discussed. 
\end{abstract}

\end{frontmatter}

%\maketitle

\section{Introduction}

The existence of a zero-point vacuum energy is one of the main tenets of the quantum formulation of the laws that we believe govern our Universe. In a Quantum Field Theory (QFT), the presence of fluctuating zero-point fields implies the existence of a non-vanishing macroscopic force between the boundaries that delimit a spatial region~\cite{Mostepanenko:1988bs}, 
due to the difference in the spectrum of quantized field modes inside and outside this region. When the boundaries of this delimited spatial domain take the form of two parallel plates, this manifestation of the vacuum fluctuation is known as the Casimir effect~\cite{Casimir:1948dh}. The computation of the Casimir force in QFT is a standard textbook exercise~\cite{zee2010quantum,radovanovic2007problem,Milton:2004ya}, and its existence, in the case of Quantum Electrodynamics, has been verified to a high precision~\cite{Lamoreaux:1996wh,Mohideen:1998iz,Roy:1999dx,Klimchitskaya:2019hon}.

The Casimir effect is now behind many experimental and theoretical pursuits. It is used as a tool to place constraints on Yukawa-type interactions~\cite{Mostepanenko:1993xy,Klimchitskaya:2015kxa}, and it has been suggested as a potential probe for the detection of feebly-interacting axion-like dark matter~\cite{Cao:2017ocv}. Casimir forces cannot be neglected at the nanoscale, and must be accounted for in the design of microelectromechanical systems~\cite{Bordag:2009zz}. Among theoretical extensions one can list its generalization to spacetimes with non-trivial topologies~\cite{DeWitt:1975ys,Ford:1975su}, dynamical boundary conditions~\cite{Juarez-Aubry:2020psk}, and non-Euclidean space-times~\cite{Dowker:1978vy,Aliev:1996va,Huang:1997br}. In the latter case it offers an independent derivation of the Hawking temperature from particle production from black holes~\cite{Nugaev:1979fj,Nugaev:1987kx}. Modifications of the Casimir effect in the presence of weak gravitational fields have been extensively studied~\cite{Sorge:2005ed,Sorge:2009zz,Bezerra:2014pza,Setare:2000py,Calloni:2001hh,Esposito:2008pn,Nazari:2015oha}. In this context, the Casimir effect can potentially provide clues on the connection between zero-point fluctuations and the cosmological constant~\cite{Caldwell:2002im,Mahajan:2006mw,Godlowski:2007gx,Szydlowski:2007bg}.

The Casimir effect stands as a potential handle to distinguish between Lorentz-invariant and Lorentz-violating formulations of QFT. Lorentz invariance (LI) is one of the cornerstones behind QFT and general relativity, and to date there are no experimental signs of a departure from it~\cite{Kostelecky:2010ux}. Nevertheless, the quantum nature of the spacetime at distances of the order of the Planck length ($\ell_P$) has been shown to provide mechanisms that can lead to violation of LI in certain formulations of quantum gravity~\cite{AmelinoCamelia:1997gz,Kostelecky:1988zi,Douglas:2001ba}. As an example, spontaneous LI breaking can occur within some string theories~\cite{Kostelecky:1988zi}. Therefore, a better understanding on the consequences of the breakdown of LI at scales larger than $\ell_P$ would provide valuable information about the microscopic structure of spacetime. In this Letter we explore the manifestation of the spontaneous breakdown of LI, induced by a constant background tensor, on the Casimir effect for a real massive scalar field between two parallel conductive plates in flat spacetime. We also explore how thermal corrections are affected by the Lorentz symmetry breaking. Our work provides a generalization of previous studies of LI violation in the Casimir context~\cite{Petrov1,Cruz:2018bqt}.

\section{The model}\label{Model}

Arguably, the most straightforward way to implement Lorentz violation is by means of the introduction of a tensor field with a non-zero vacuum expectation value  (VEV). When coupled to the Standard Model fields the spontaneous symmetry breaking, induced by the non-zero VEV, is manifested as preferential directions on the spacetime, leading to a breakdown of LI. In the case of a real scalar field in flat spacetime, with the Minkowski metric with signature $(+,-,-,-)$, we parametrize this coupling in the following form,
\begin{align}\label{mod KG Lagrangian}
\mathcal{L} = \frac{1}{2} h^{\,\mu\nu}\partial _{\mu} \phi\,\partial _{\nu} \phi - \frac{1}{2} m ^{2} \phi ^{2}\,.
\end{align}
Here $h^{\,\mu\nu}$ is a symmetric tensor that represents a constant background, independent of the spacetime position, and which does not transform as a second order tensor under {\em active} Lorentz transformations\footnote{Single derivative terms, such as $i\phi u^\mu\partial_\mu \phi$ with $u^\mu$ a constant 4-vector, can be reduced to surface terms, which in absence of topological effects do not have physical contributions~\cite{KosteleckyEscalar}.}. Naturally, causality, the positive energy condition and stability impose restrictions on the components of $h^{\, \mu \nu}$. 

\par\medskip

Consider now the following set-up: a pair of parallel, conductive plates, orthogonal to the $\hat{z}$-direction, located at $z=0$ and $z=L$, on which Dirichlet boundary conditions apply for the field $\phi$. That is, $\phi(z=0)=\phi(z=L)=0$. We now solve for the scalar field between the plates, applying the Green's function technique~\cite{milton2001the}. Namely, we are interested in computing the time-ordered, vacuum two-point correlation function $G(x,x')=-i\langle 0| \mathcal{T} \phi(x) \phi(x')|0\rangle$, which as is well known (see e.g.~\cite{Peskin:1995ev}) satisfies the Green's function (GF) equation
\begin{align}\label{EQG2a}
\mathcal{O}_{\vec{x}}\,G (x,x^{\prime}) = \delta^{(4)} (x - x ^{\prime}).
\end{align}
Here, in the configuration space, the modified Klein-Gordon operator has the following explicit form,
\begin{equation}
\mathcal{O}_{\vec{x}}=h^{00}\partial_0^2 + 2h^{0\bar{i}}\partial _{0}\partial _{\,\bar{i}}+h^{\bar{i}\bar{j}}\partial _{\,\bar{i}}\partial _{\,\bar{j}}+2h^{03}\partial _{0}\partial _{3}+2h^{\bar{i}3}\partial _{\,\bar{i}}\partial _{3} +h^{33}\partial^2 _{z}+m^2
\end{equation}
with $\bar{i},\bar{j}=1,2$. In the chosen coordinate system, the GF is invariant under translations in the $(\hat{x},\hat{y})$-plane. Taking advantage of this symmetry, we can express the GF in terms of the Fourier transform in the direction parallel to the plates, 
\begin{equation}\label{GFGeneral}
G(x,x')=\int\frac{d^2 \vec{k}_\bot}{(2\pi)^2}  e^{i\vec{k}_\bot\cdot(\vec{x}_\bot-\vec{x}_\bot')}\int \frac{d\omega}{2\pi}e^{-i\omega(t-t')}g\left(z,z';\omega,\vec{k}_\bot\right),
\end{equation}
where $\vec{k}_\bot=(k_x,k_y)$ and $\vec{x}_\bot=(x,y)$. Henceforth we will drop the explicit dependence on $\omega$, $\vec{k}_\bot$ of $g$ for simplicity. After substitution of (\ref{GFGeneral}) into (\ref{EQG2a}), and straightforward integration of the resulting 1D boundary problem,\footnote{An analogous step-by-step procedure can be found in~\cite{milton2001the}.} an exact solution for the reduced GF between the plates can be found,
\begin{align}
g _{\parallel} (z,z ^{\prime}) \;&=\; e ^{- i \xi _{0} (z ^{\prime} - z)} \frac{\sin ( \xi _{1} z _{<}) \sin [ \xi _{1} (z _{>} -L) ]}{ h^{33} \xi _{1} \sin (\xi _{1} L) }\,. \label{gz1}
\end{align}
Here $z _{>}$ ($z _{<}$) is the greater (lesser) between $z$ and $z ^{\prime}$. The coefficients $\xi_{0,1}$ denote the following combinations of energy-momenta and the Lorentz-violating tensor, 
\begin{align}
\xi_0 \;&=\; \frac{1}{h^{33}}(h^{03}\omega-h^{\bar{i}3}\vec{k}_{\bot\bar{i}})\,,\\ %\notag
\xi_1 \;&=\;\frac{1}{|h^{33}|} \left[ (h^{03}\omega-h^{\bar{i}3}\vec{k}_{\bot\bar{i}})^2 - h^{33}  \gamma^2  \right]^{1/2}\,,
\end{align}
where
\begin{equation}
\gamma ^{2} \;=\; h^{00}\omega ^{2} - 2h^{0\bar{i}}\omega k _{\perp_{\bar{i}}} + h^{\bar{i}\bar{j}}\vec{k}_{\perp_{\bar{i}}}\vec{k}_{\perp_{\bar{j}}}-m^2\,.
\end{equation}
The LI limit is recovered by taking $h ^{\, \mu \nu} \to \eta ^{\, \mu \nu}$, which implies $\xi_0\rightarrow 0$ and $\xi_1^2\rightarrow \omega ^{2} - k _{\perp} ^{2} - m ^{2}$. It is worth noting that the case with Neumann conditions can be trivially recovered by replacing $\sin\rightarrow\cos$ in the numerator of (\ref{gz1}).

The determination of the Casimir energy and stress requires not only the GF for the two plate setup, but also the GF in the presence of no plates and a single plate. For the former, we find
\begin{align}
g _{v} (z,z ^{\prime}) \;&=\; - \frac{i}{2\xi_1} \frac{e^{i\xi_0(z-z')}}{h^{33}} e^{i\xi_1 (z_> - z<)} \,, \label{gz2}
\end{align}
while for the latter,
\begin{align}
g _{|} (z,z ^{\prime}) \;&=\; -\frac{e^{i\xi_0(z-z')}}{\xi_1h^{33}} \sin[\xi_1(z_<-L)] e^{i\xi_1 (z_> - L)} \,. \label{gz3}
\end{align}

\par\medskip

In order to quantify the Casimir effect, we need an expression for the vacuum expectation value for the stress-energy tensor of the scalar field, $T ^{\mu \nu} = h ^{\,\mu\alpha}\partial_\alpha \phi\,\partial^\nu \phi -\eta ^{\,\mu \nu}\mathcal{L}$. In terms of the GF, it can be generically computed as~\cite{milton2001the}
\begin{equation}\label{VEVTmunu}
\langle T^{\mu\nu}\rangle = -i\lim_{x\rightarrow x^\prime}\left[h^{\mu\alpha}\partial_\alpha\,\partial^{'\nu}\right]G(x,x^\prime) - \eta^{\,\mu\nu} \langle\mathcal{L}\rangle \; ,
\end{equation}
while the VEV of the Lagrangian density can be written as $\langle\mathcal{L}\rangle = -i \lim_{x\rightarrow x^\prime}\frac{1}{2}\left( h^{\,\mu\nu}\partial_\mu\,\partial_{'\nu}-m^2\right) G(x,x^\prime)$.  \par\medskip

Substitution of (\ref{GFGeneral}) leads to the following expressions for the energy density and the pressure in the $\hat{z}$-direction,
\begin{align}
\langle T^{00}\rangle \;=\; &- i \lim _{z ^{\prime} \to z} \int \frac{d \omega}{2 \pi} \int \frac{d ^{2} \vec{k} _{\perp}}{(2 \pi) ^{2}}\left[h^{00} \omega ^{2} - h^{0\bar{i}}\omega \vec{k}_{\bot_{\bar{i}}}\right.\nonumber\\
&\left.+ih^{03}\omega\partial_z\right] g (z,z ^{\prime}) - \langle \mathcal{L} \rangle , \label{T00a}\\ %\notag
\langle T^{33} \rangle \;=\; &- \frac{i}{2} \lim _{z ^{\prime} \to z} \int \frac{d \omega}{2 \pi} \int \frac{d ^{2} \vec{k} _{\perp}}{(2 \pi) ^{2}} \left[ \gamma ^{2} -  h^{33} \partial _{z} \partial _{z ^{\prime}} \right]  g (z , z ^{\prime} ) . \label{ExpValTzz}
\end{align}
where 
\begin{align} \notag
\langle\mathcal{L}\rangle 
\;=\; &- \frac{i}{2} \lim _{z ^{\prime} \to z} \int \frac{d \omega}{2 \pi} \int \frac{d ^{2} \vec{k} _{\perp}}{(2 \pi) ^{2}} \bigg[ \gamma ^{2} + h^{33} \partial _{z} \partial _{z ^{\prime}} \\
&-ih^{33}\xi_0(\partial_{z^\prime}-\partial_z) \bigg] \;  g (z , z ^{\prime} )  \; . \label{ExpValL}
\end{align}

%%%%%%%%%%%%%%%%%%%%%%%%%%%%%%%%%%%%%%%%%%%%%%
%%%%%%%%%%%%%%%%%%%%%%%%%%%%%%%%%%%%%%%%%%%%%%
%%%%%%%%%%%%%%%%%%%%%%%%%%%%%%%%%%%%%%%%%%%%%%

\section{Casimir Effect with Lorentz symmetry violation} \label{CEsection}

With the VEV of the stress-energy tensor at hand, we now proceed to compute the global Casimir energy and the Casimir stress upon the plates in the presence of Lorentz-invariance violation.

\subsection{Global Casimir energy} \label{GlobalCasSec}

The renormalized vacuum energy stored between the parallel plates can be computed  formally as the difference between the zero-point energy in the presence of the boundary, $\langle T^{00} \rangle _{\parallel}$, and that of the free vacuum, $\langle T^{00} \rangle_{v}$. Namely,
\begin{equation}
    \mathcal{E} _{C} (L) = \int _{0} ^{L} \left( \langle T^{00} \rangle _{\parallel} - \langle T^{00} \rangle _{v} \right) \, dz\,.
    \label{Eglobal}
\end{equation}
We begin by evaluating $\langle T^{00} \rangle _{\parallel}$. As a first step, it can be noted after a cursory computation that the contribution from the VEV of $\mathcal L$ in (\ref{T00a}) is $L$-independent and will therefore not contribute to the Casimir pressure.\footnote{More precisely, $\int _{0} ^{L} \langle \mathcal{L} \rangle _{\parallel} \, dz = (1/2i) \int \frac{d \omega}{2 \pi} \int \frac{d ^{2} \vec{k} _{\perp}}{(2 \pi) ^{2}}$.} After simplification, the remaining terms in (\ref{T00a}) can be rearranged to lead to the following expression,
\begin{align}\notag
\langle T^{00} \rangle_{\parallel} = - i \int \frac{d \omega}{2 \pi} \int \frac{d ^{2} \vec{k} _{\perp}}{(2 \pi) ^{2}} \Big[h^{00} \omega ^{2} &-h^{0\bar{i}}\omega (\vec{k}_\bot)_{\,\bar{i}} \\
&  -h^{03}\omega\,\xi_0  \Big] g _{\parallel} (z,z)\,. \label{T00Plates}
\end{align}
The term inside the brackets in the previous equation is a quadratic form in $(\omega,k _{x} , k _{y})$, with coefficients given by the components of $h^{\,\mu\nu}$. This quadratic form is different from that appearing in the argument of the GF, $|\,h^{33}|\,\xi_1^2$, and this makes the evaluation of (\ref{T00Plates}) a non-trivial task. However, a closed-form solution may be obtained by diagonalization of the latter quadratic form, mapping it into a mimic of the LI case, $|\,h^{33}|\,\xi_1^2=\omega ^{ \prime \, 2} - k _{x} ^{\prime \, 2} - k _{y} ^{\prime \, 2} - m ^{2}$, where primed quantities correspond to the rotated frequency and momenta. Further performing a Wick rotation $\omega' \rightarrow i\zeta$, it can be shown that (\ref{T00Plates}) is equivalent to the following expression,
\begin{align}
\langle T^{00} \rangle_{\parallel} &= \frac{1}{\sqrt{-h}} \int \frac{d \zeta}{2 \pi} \int \frac{d ^{2} \vec{k^{\prime}} _{\perp}}{(2 \pi) ^{2}} \, \zeta ^{2} \, \frac{\sinh ( \gamma \tilde{z} ) \sinh [ \gamma (\tilde{z}-\tilde{L}) ]}{\gamma \sinh (\gamma \tilde{L}) } , \label{T00Plates4}
\end{align}
where now $\gamma ^{2} = \zeta ^{2} + k _{\perp} ^{\prime \, 2} + m ^{2}$, $h\equiv \textrm{det}\,h^{\,\mu\nu}$, and 
\begin{equation}\label{rescaledz}
\tilde{z} =\frac{ z}{ \sqrt{-h^{33}}}\,, \quad \tilde{L} = \frac{L}{ \sqrt{-h^{33}} }\,.
\end{equation}

An entirely analogous procedure can be followed to evaluate the vacuum energy density $\langle T^{00} \rangle _{v}$, making use in this case of the  corresponding GF (\ref{gz2}). For it we obtain
\begin{align}
\langle T^{00} \rangle _{v} &= - \frac{1}{\sqrt{-h}} \int \frac{d \zeta}{2 \pi} \int \frac{d ^{2} \vec{k}' _{\perp}}{(2 \pi) ^{2}} \, \frac{\zeta ^{2}}{2 \gamma} . \label{T00Vacuum}
\end{align}
Finally, substituting into (\ref{Eglobal}), integrating with respect to $z$ and dropping an $L$-independent constant term leads to the following expression for the vacuum energy between the plates,
\begin{align}
    \mathcal{E} _{C} (L) = - \sqrt{\frac{h^{33}}{h}} \int \frac{d \zeta}{2 \pi} \int \frac{d ^{2} \vec{k}'_{\perp}}{(2 \pi) ^{2}} \frac{\zeta ^{2}}{2 \gamma} \tilde{L} \, [ \coth{(\gamma \tilde{L})} - 1 ] .
     \label{EFinal2a}
\end{align}
This resulting integral can be recognized as the LI result, $\mathcal{E} _{0}$, rescaled by the factor $\sqrt{h^{33}/h}$, with a rescaled separation between the plates (\ref{rescaledz})~\cite{milton2001the}. Integration gives
\begin{align}
    \mathcal{E} _{C} (L)= \sqrt{\frac{h^{33}}{h}} \mathcal{E} _{0} (\tilde{L})= -  \dfrac{m ^{2}}{8 \pi ^{2} \tilde{L}} \sqrt{\frac{h^{33}}{h}}  {\displaystyle \sum} _{n = 1} ^{\infty} \dfrac{1}{n ^{2}} K _{2} (2 m n \tilde{L}), \label{CasimirEnergyFin}
\end{align}
where $K _{2} (x)$ is the second-order Bessel function of the second kind. Note that the Lorentz-violating result reduces trivially to the LI one as $h^{33}, h \rightarrow -1$, which would be the case for $h^{\mu\nu}\rightarrow \eta^{\mu\nu}$. Although the sum which appears in (\ref{CasimirEnergyFin}) does not have an analytical closed form, it can be reduced to simple expressions in the large and small mass limits,
\begin{align}
\mathcal{E} _{C} (L) \;\simeq\; - \sqrt{\frac{h^{33}}{h}} \times \begin{cases} \dfrac{\pi ^{2}}{1440 \tilde{L} ^{3}} - \dfrac{m ^{2}}{96 \tilde{L}} \,, & m\tilde{L}\ll 1\,,\\[8pt] \dfrac{m ^{2}}{16 \pi ^{2} \tilde{L}} \sqrt{\dfrac{\pi}{m \tilde{L}}} e ^{- 2m \tilde{L}} \,, & m\tilde{L}\gg 1\,. \end{cases}
\end{align}
The massless case is trivially recovered taking the $m\rightarrow0$ limit in the previous equation.

\subsection{Stress on the plates} \label{ST}

We now proceed to determine the Casimir stress upon the plate at $z = L$ by direct evaluation of the normal-normal component of the stress-energy tensor (\ref{ExpValTzz}). Denoting by $\langle T ^{33} \rangle _{\parallel}$ the vacuum stress due to the confined scalar field, and by $\langle T ^{33} \rangle _{\vert}$ the stress due to the field above the plate, we can write
 \begin{align}
    \mathcal{F} _{C} (L) = \langle T ^{33} \rangle_{\parallel} - \langle T ^{33} \rangle _{\vert} . \label{CasimirStress}
\end{align} 
In a similar fashion to the previous computation of the Casimir energy, all it takes to calculate these stresses is to substitute the corresponding reduced GFs into (\ref{ExpValTzz}), and to repeat the quadratic form diagonalization procedure. A key difference in this analysis is the fact that the Lagrangian density {\em does} contribute to the stress.\footnote{ $\langle \mathcal{L}\rangle$ also plays a fundamental role regarding the behavior of the field near the boundaries, see Section~\ref{LE}.} Nevertheless, despite this relative complication, a  straightforward calculation using (\ref{gz1}) and (\ref{gz3}) yields
\begin{equation}
\begin{aligned}
\label{T331}
\langle T^{33} \rangle_{\parallel} \;&=\; - \frac{1}{\sqrt{-h}} \int \frac{d \zeta}{2 \pi} \int \frac{d ^{2} \vec{k}' _{\perp}}{(2 \pi) ^{2}}  \frac{\gamma}{2}  \coth (\gamma \tilde{L})\,,\\
\langle T^{33} \rangle_{\vert} \;&=\; - \frac{1}{\sqrt{-h}} \int \frac{d \zeta}{2 \pi} \int \frac{d ^{2} \vec{k}'_{\perp}}{(2 \pi) ^{2}}  \frac{\gamma}{2}\, . 
\end{aligned}
\end{equation}
Each stress contains an $L$-independent divergent term that is canceled by the regularization provided by (\ref{CasimirStress}). Substitution of these expressions into (\ref{CasimirStress}), and following the same steps that lead to Eq.~(\ref{EFinal2a}), produces
\begin{align}
    \mathcal{F} _{C} (L)&= \frac{1}{\sqrt{-h}} \mathcal{F} _{0} (\tilde{L})=  \frac{1}{\sqrt{-h}} \dfrac{1}{4 \pi ^{2}} {\displaystyle \int _{0} ^{\infty}} \dfrac{\tau ^{2} \sqrt{\tau ^{2} + m ^{2}}}{e ^{2\tilde{L} \sqrt{\tau ^{2} + m ^{2}}} - 1} d \tau. \label{CasimirForceFin}
    \end{align}
In this expression $\mathcal{F} _{0}$ denotes the LI result. Expectedly, the stress in the LI violating result is proportional to the stress in the absence of Lorentz violation, but evaluated at the rescaled length $\tilde{L}$. For a vanishing scalar field mass, Eq.~(\ref{CasimirForceFin}) reduces to $ \mathcal{F} _{C} (L)|_{m=0}=- \pi^2/(480 \tilde{L}^4\sqrt{-h})$.

As a consistency check, one can verify that the Casimir energy (\ref{CasimirEnergyFin}) and the stress (\ref{CasimirForceFin}) are connected by the elementary relation
\begin{align}
    \mathcal{F} _{C} (L) = - \frac{\partial \mathcal{E}_{C} (L)}{\partial L} .
\end{align}

\subsection{Local effects} \label{LE}

In Section~\ref{GlobalCasSec} we derived an expression for the global Casimir energy by computing the integral of $\langle T^{00} \rangle _{\parallel} - \langle T^{00} \rangle _{v}$ in the region between the plates by means of the GF method. Although alternative methods exist to evaluate $\mathcal{E}_C$~\cite{Plunien:1986ca}, the power of the GF procedure arises clearly when studying the {\em local} energy density, which in turn reveals the divergence structure of the theory. The computation of $\langle T^{\mu\nu}\rangle$ is the goal of this section.

We begin with the energy density per unit volume between the plates. Without dropping in this case the contribution of $\langle \mathcal{L}\rangle$ (which was discarded in the global analysis due to its $L$-independence after integration), the same analysis that led to (\ref{T00Plates4}) in this case gives
\begin{align} \notag
     \langle T^{00} \rangle \;=\; &- \frac{1}{\sqrt{-h}} \int \frac{d \zeta}{2 \pi} \int \frac{d ^{2} \vec{k}'_{\perp}}{(2 \pi) ^{2}} \\
     &\times\left\lbrace \frac{\zeta ^{2}}{2 \gamma} \coth (\gamma \tilde{L}) + \frac{k_{\perp} ^{\prime\,2} + m ^{2}}{2 \gamma} \frac{\cosh [\gamma (2 \tilde{z} - \tilde{L})]}{\sinh (\gamma \tilde{L})} \right\rbrace .
\end{align}
The introduction of the polar coordinates $k _{\perp} = \rho \cos \theta$, $\zeta = \rho \sin \theta$, where $\rho \in [0, \infty)$ and $\theta \in [- \pi / 2 , \pi / 2]$, leads to the following result
\begin{eqnarray}
    \langle T^{00} \rangle &=& - \frac{1}{12 \pi ^{2}} \frac{1}{\sqrt{-h}} \int _{0} ^{\infty} \left\lbrace \frac{\rho ^{4}}{\gamma ^{\ast}} \frac{2}{e ^{2 \gamma ^{\ast} \tilde{L}} - 1} \right. \nonumber\\
    &&\left. + \frac{\rho ^{2}}{\gamma ^{\ast}} (2 \gamma ^{\ast \, 2} + m ^{2}) \frac{e ^{2 \gamma ^{\ast} \tilde{z}} + e ^{2 \gamma ^{\ast} ( \tilde{L} - \tilde{z} )}}{e ^{2 \gamma ^{\ast} \tilde{L}} - 1} \right\rbrace d \rho\,.  \label{LocalEffects}
\end{eqnarray}
Here $\gamma ^{\ast} = \sqrt{\rho ^{2} + m ^{2}}$, and we have discarded an $L$-independent term. Denoting by $\mathfrak{U}$ the $z$-independent term in the previous expression, one can easily show that $\mathfrak{U}=\mathcal{E}_C/L$. Similarly, a straightforward change of variables allows us to write the $z$-dependent term of (\ref{LocalEffects}), which we denote by $f(z)$, as follows,
\begin{eqnarray}
    f (z) &=& - \frac{1}{192 \pi ^{2} \tilde{L} ^{4}} \frac{1}{\sqrt{-h}} \int _{2 m \tilde{L}} ^{\infty} \sqrt{y ^{2} - (2 m \tilde{L}) ^{2}} \nonumber\\
    &&\times\left[ 2y ^{2} + (2 m \tilde{L}) ^{2} \right] \frac{e ^{y z/L} + e ^{y(1-z/L)}}{e ^{y} -1} dy . \label{fz}
\end{eqnarray}
In the massless limit, this function can be expressed in terms of the Hurwirtz zeta function, $\zeta (s,a) = \sum _{n = 0} ^{\infty} (n+a) ^{-s}$, 
\begin{align}
    f (z) \;&=\; - \frac{1}{16 \pi ^{2} L ^{4}} \frac{(h^{33}) ^{2}}{\sqrt{-h}} \left[ \zeta (4, z/L) + \zeta (4, 1 - z/L) \right]\,. \label{fz-massless}
\end{align}

Therefore we have found that $ \langle T^{00} \rangle = \mathfrak{U} + f (z)$. $\mathfrak{U}$ encodes the part of the vacuum energy resulting in an observable force, whereas $f (z)$ corresponds to a local, divergent effect that does not contribute to the pressure, as the $L$-independence of the following integral confirms
\begin{align}
    \int _{0} ^{L} f(z) dz = - \frac{1}{48 \pi ^{2}} \sqrt{\frac{h^{33}}{h}} \int _{2 m} ^{\infty} \sqrt{x ^{2} - 4 m ^{2}} ( x ^{2} + 2 m ^{2} ) \frac{dx}{x}.
\end{align}
In the massless case this divergence is quartic as $z$ approaches the plates, as can be appreciated from Eq.~(\ref{fz-massless}). For a generic mass the complex form of (\ref{fz}) prevents us from analytically determining the degree of divergence. 

We turn now to the evaluation of the VEV for the remaining components of $T^{\mu\nu}$. Owing to the symmetry of the setup, these components can be easily determined. For example, rotational invariance around the $z$-axis immediately implies that $\langle T^{11} \rangle=\langle T^{22} \rangle$. Moreover, after an explicit calculation we find that $\langle T^{11} \rangle = - \langle T^{00} \rangle$. The off-diagonal components of $T^{\mu\nu}$ vanish in the LI limit, but in the presence of a non-trivial $h^{\,\mu\nu}$ they are in general non-zero, although they can also be related to the $00$ and $33$ components by symmetry arguments. A cursory computation provides the following general expression for the VEV of the stress-energy tensor,
\begin{eqnarray}
\langle T^{\mu \nu} \rangle &=& - \frac{2 h^{\alpha 3}}{h^{33}} (\eta ^{\,\mu \alpha} + n ^{\mu} n ^{\alpha})  n ^{\nu}  \left[\langle T^{00} \rangle - \langle T^{33} \rangle - f(z) \right]  \nonumber\\
&& + (\eta ^{\,\mu \nu} + n ^{\mu} n ^{\nu}) \, \langle T^{00}\rangle +  n ^{\mu} n ^{\nu} \, \langle T^{33} \rangle \,. \label{VSC}
\end{eqnarray}
Here $n ^{\mu} = (0,0,0,1)$ is the unit vector perpendicular to the plates, and $\langle T^{00} \rangle$ and $\langle T^{33} \rangle$ are given by Eqs.~(\ref{LocalEffects}) and~(\ref{T331}), respectively. Clearly, in the LI limit the first term vanishes and we recover the usual structure of the vacuum stress~\cite{Brown:1969na}.

\subsection{Finite temperature effects}

The Casimir effect, as described in the previous sections, is a manifestation of the fluctuations of the $\phi$ field in the {\em vacuum}. However, any realistic parallel plate setup will necessarily be immersed in a bath with a temperature above absolute zero. It is therefore crucial to determine the effect that {\em thermal} fluctuations would have in the Casimir stress. Luckily, in our relatively simple scenario, the stress at $T>0$ case can be determined in a straightforward manner.

In the Matsubara formalism of finite temperature QFT, the Casimir stress at nonzero temperature can be obtained from Eq.~(\ref{T331}) upon the replacement $\int d\zeta/2\pi \rightarrow \beta^{-1} \sum_{n=-\infty}^{\infty}$, together with mapping the imaginary frequency $\zeta$ to the discrete Matsubara frequency $\zeta_n \equiv 2\pi n/\beta$~\cite{Kapusta:2006pm}. Here $\beta = 1 / k _{B} T$, with $k _{B}$ the Boltzmann constant. These substitutions yield
\begin{align}
    \mathcal{F} _{C} (L;T) = - \frac{1}{\beta \sqrt{h}}
 \sum _{n = - \infty} ^{+ \infty} \int \frac{d ^{2}
\vec{k} _{\perp}}{(2 \pi) ^{2}} \frac{\gamma _{n}}{e ^{2 \gamma _{n}
\tilde{L}} -1}\, , \label{FiniteTemp}
\end{align}
where $\gamma _{n} = \sqrt{\zeta _{n} ^{2} + k _{\perp} ^{2} + m ^{2}}$. Although this expression lacks a closed form in terms of elementary functions, we can gain some insight of its behavior in the massless case for small temperature and large temperature (classical) limits. For low temperature, the above expression for the pressure takes the form
\begin{align}
\mathcal{F} _{C} (L;T \ll 1) \approx - \frac{\pi ^{2}}{480 \tilde{L} ^{4} \sqrt{h}} \left( 1  + \frac{1}{48 \pi ^{4}} s ^{4} - \frac{60}{\pi ^{2}} s e ^{- 4 \pi ^{2} / s} \right) , \label{LowT}
\end{align}
where $s = 4 \pi k _{B} T \tilde{L} \ll 1$. Clearly this result is consistent with the Nernst heat theorem, since the associated entropy vanishes as $s$ goes to zero.

In the opposite regime, at high temperatures, all terms in the sum of Eq. (\ref{FiniteTemp}) except the $n=0$ term are exponentially suppressed, resulting in 
\begin{align}
    \mathcal{F} _{C} (L;T\gg 1) \approx - \frac{\zeta (3) k _{B} T}{8 \pi \tilde{L} ^{3} \sqrt{h}} - \frac{k _{B} T}{4 \pi \tilde{L} ^{3} \sqrt{h}} \left( 1 + s + \frac{s ^{2}}{2} \right) e ^{-s}  ,  \label{HighT}
\end{align}
where here $s \gg 1$. The leading term can also be obtained from the Helmholtz free energy for Lorentz-violating massless bosons.

The results of equations  (\ref{LowT}) and (\ref{HighT}) exhibit an interesting behaviour as a function of the Lorentz violating parameter $h ^{33}$ through the rescaled length $\tilde{L} = L / \sqrt{- h ^{33}}$. When Lorentz invariance is mildly broken, $h ^{33} \approx - 1$, and hence the conditions $s \ll 1$ and $s\gg 1$ correspond to low and high temperatures, respectively. However, when Lorentz symmetry breaking is not negligible, such conditions are relaxed and possibly flipped. For example, when $h ^{33} \approx 0 ^{-}$, the condition $s \gg 1$ can be fulfilled even for low temperatures.

\section{Summary and discussion} \label{Conclusection}

In the present work we have obtained explicit expressions for the Casimir energy and force between two parallel conductive plates, arising from the vacuum fluctuations of a massive real scalar field, in the presence of a generic background defined by the tensor $h^{\,\mu\nu}$ in Eq.~(\ref{mod KG Lagrangian}). This background is motivated by theories in which the breakdown of Lorentz invariance manifests itself as the non-vanishing vacuum expectation value of a fundamental field.

Since no deviation from Lorentz invariance has been experimentally observed yet, the perturbative expansion $h^{\,\mu\nu}=\eta^{\,\mu\nu}+k^{\mu\nu}$ is justified. Here $\eta^{\,\mu\nu}$ is the Minkowski metric and $k^{\mu\nu}$ is a constant tensor whose components are much smaller than one $\vert k^{\, \mu\nu}\vert \ll1$. Working to first order in $k^{\, \mu\nu}$, it is possible to prove that the Lorentz-violating theory described by Eq.~(\ref{mod KG Lagrangian}) can be transformed into the standard Lorentz-invariant theory by an appropriate change of spacetime coordinates $x^{\prime \, \mu}=x ^{\, \mu} - \frac{1}{2}k\, ^\mu\,_\nu x^{\nu}$ \cite{Ferrero:2011yu}. In this new coordinate system it is relatively straightforward to evaluate the Casimir energy. It is given by the Lorentz-invariant result, albeit with a redefinition of the separation between the plates and a global multiplicative factor arising from the Jacobian of the transformation. Let us discuss our result in Eq.~(\ref{CasimirForceFin}) in this approximation. One can verify that the global multiplicative factor, $1/\sqrt{-h}$ in Eq.~(\ref{CasimirForceFin}), corresponds to the square root of the Jacobian, whereas $\tilde{L}\approx L(1+\frac{1}{2}k^{33})$ is precisely the transformed distance between plates. This confirms that our result, valid to all orders in $k^{\mu\nu}$, correctly reduces to the expected result in the limit $\vert k^{\mu\nu}\vert\ll1$.

Focusing on the massless case for simplicity, the (measurable) Casimir force explicitly reduces to first order in $k^{\mu\nu}$ to $\mathcal{F}_C(L)=(1-2k^{33} - \frac{1}{2} \eta ^{\, \mu \nu} k _{\mu \nu}) \, \mathcal{F}_0(L)$. For the sake of comparison, if we consider the present experimental measurements of the Casimir force between parallel plates for the electromagnetic case (15\% precision in the 0.5-3 $\mu$m range), the bound that can be obtained from this result is $|\, 2k^{33} + \frac{1}{2} \eta ^{\, \mu \nu} k _{\mu \nu}|<10^{-2}$. Note that the leading-order modification to the Lorentz invariant result only involves the component of $k^{\mu\nu}$ perpendicular to the plates and the trace of $k^{\mu\nu}$. We also note that in this Letter we have assumed that Dirichlet boundary conditions apply at the plates location. Nevertheless, other types of boundary conditions, such as Neumann conditions, can be treated in a completely analogous manner since they only directly modify the Green's function form. We have found that for this parallel plate setup, the form of the Casimir energy and force are independent of the choice of Dirichlet or Neumann conditions, as happens in the LI case.

It is worth mentioning that in Refs.~\cite{Petrov1,Cruz:2018bqt} the Casimir effect and its corresponding thermal corrections for the scalar field were studied for a particular case where $h^{\,\mu\nu}=\eta^{\,\mu\nu}+\lambda u\,^\mu u\,^\nu$, being $\lambda$ a LV parameter and $u\,^\mu$ a four-vector that specifies the direction in which the Lorentz symmetry is broken. There, the authors considered separately different choices of the four-vector $u\,^\mu$ and analyzed, by means of the mode-summation method, the Casimir effect.  One can verify that our results in Eqs.~(\ref{CasimirEnergyFin}) and (\ref{FiniteTemp}) for the global Casimir energy and thermal corrections to the Casimir stress respectively reduce to the ones reported in Refs.~\cite{Petrov1,Cruz:2018bqt} by setting $h^{\,\mu\nu}=\eta^{\,\mu\nu}+\lambda u\,^\mu u\,^\nu$. However, the local approach adopted here provides additional information regarding the local behavior of the theory, besides the generalization and flexibility that the second-rank tensor $h ^{\, \mu \nu}$ gives to the model.

We finish by emphasizing that our method allowed us to determine the effect of $h^{\,\mu\nu}$ on the Casimir energy and stress in a {\em non-perturbative} way and did not require a smallness condition on the magnitude of the components of $h^{\,\mu\nu}$. Although this appears to be an overkill in the context of Lorentz invariance violation, our computation can be relevant for condensed matter physics and materials science because therein the internal structure of media, which generically leads to anisotropies, will play an analogous role to that of a background in empty space.

\section{Acknowledgements}

A. M.-R. acknowledges support from DGAPA-UNAM Project No. IA101320. C. A. E. is supported by a UNAM- DGAPA postdoctoral fellowship and Project PAPIIT No. IN111518. M. A. G. G. is supported by the Spanish Agencia Estatal de Investigaci\'on through the grants FPA2015-65929-P (MINECO/FEDER, UE), PGC2018095161-B-I00, IFT Centro de Excelencia Severo Ochoa SEV-2016-0597, and Red Consolider MultiDark FPA2017-90566-REDC. O. J. F. acknowledges support from DGAPA-UNAM Project No. IN103319.

\bibliographystyle{elsarticle-num}

\bibliography{biblio}

\end{document}